\documentclass[12pt]{iopart}

\usepackage{iopams}
\usepackage{units}
\usepackage{graphicx}
\usepackage[rgb]{xcolor}
\usepackage[square,comma,numbers,compress]{natbib}

\makeatletter
\def\footnoterule{\kern-3\p@
	\hrule \@width 2in \kern 2.6\p@} 
\@definecounter{footnote}
\def\thefootnote{\@arabic\c@footnote}
\makeatother

\begin{document}

\title[Micromechanical gravitation between small objects]{A micromechanical proof-of-principle experiment for measuring the gravitational force of milligram masses}

\author{Jonas Schm\"ole, Mathias Dragosits, Hans Hepach, Markus Aspelmeyer}

\address{Vienna Center for Quantum Science and Technology (VCQ), Faculty of Physics, University of Vienna, Boltzmanngasse 5, Vienna, Austria}
\ead{jonas.schmoele@univie.ac.at, markus.aspelmeyer@univie.ac.at}
\vspace{10pt}
\begin{indented}
\item[]\today
\end{indented}

\begin{abstract}
This paper addresses a simple question: how small can one make a gravitational source mass and still detect its gravitational coupling to a nearby test mass? We describe an experimental scheme based on micromechanical sensing to observe gravity between milligram-scale source masses, thereby improving the current smallest source mass values by three orders of magnitude and possibly even more. We also discuss the implications of such measurements both for improved precision measurements of Newton's constant and for a new generation of experiments at the interface between quantum physics and gravity.
\end{abstract}

\section{Introduction}
Measuring gravitational forces between non-celestial bodies started with the pioneering experiments of Maskelyne \cite{Maskelyne1775AnAccount} and Cavendish \cite{Cavendish1798Density} and has remained a challenging task ever since. In astronomical observations the gravity of extremely large masses dominates their dynamics and allows confirmation of the predictions of general relativity \cite{Will2006TheConfrontation}, our best working theory of gravity, with striking agreement. The recent direct observation of gravitational waves, a consequence of general relativity, opens up an entire realm of astronomical objects which now, for the first time, become accessible to our observation \cite{Abbott2016Observation}. For masses on laboratory scales, however, the gravitational force is minuscule, making it difficult to observe effects generated by small objects. Nevertheless, Earth-based laboratory experiments have been able to achieve high-precision tests of gravity involving source masses, i.e. objects generating a gravitational field, that are typically on the order of several $\unit{kg}$ and heavier \cite{Speake2014TheSearch,Gillies2015AttractingMasses}. 

To date, the smallest source mass that has been used to produce a measurable gravitational force is around $90\unit{g}$ in the form of two $20\unit{mm}$ diameter Dy-Fe cylinders \cite{Ritter1990Equivalence} used in a torsional pendulum configuration. Here we address the question of how to measure the gravitational field of significantly smaller source masses. Its answer is enabled by the development of micro-mechanical devices, which over the last decades has resulted in sensors with unprecedented sensitivity. Examples include mechanical measurements of single electron spins \cite{Rugar2004SingleSpin,Kolkowitz2012CoherentSensing}, of superconducting persistent currents \cite{BleszynskiJayich2009PersistentCurrents} or of quantum mechanical photon fluctuations (shot noise) in a laser beam \cite{Purdy2013Observation}. In their most simplified version, such mechanical force sensors resemble an harmonic oscillator of quality factor $Q=\omega_0/\gamma$ ($\omega_0$: mechanical resonance frequency; $\gamma$: viscous damping rate) that is coupled to a thermal bath at temperature~$T$ and driven on or near its mechanical resonance by an external force. The main limitation on the sensing performance is due to thermally induced amplitude fluctuations that scale with the thermal energy $k_BT$ ($k_B$: Boltzmann's constant), the mechanical damping $\gamma$ and the resonator mass $m$. Over a certain measurement time $\tau$ this accumulates to a Brownian force noise of amplitude $F_{th}=(k_B T m \gamma /\tau)^{1/2}$ \cite{Mamin2001SubAttonewton}, which sets a lower limit for the detection of external forces. As a consequence, high-$Q$ nano-mechanical oscillators at low temperatures have already reported force sensitivities on the zepto-Newton scale \cite{Biercuk2010UltrasensitiveDetection,Moser2013Ultrasensitive,Schreppler2014QuantumMetrology}. This opens up the possibility to measure small gravitational forces. For example, let us consider a spherical mass, say a $1\unit{mm}$ radius lead sphere ($m\approx 40\unit{mg}$), trapped harmonically at a frequency of $\omega_m=100\unit{Hz}$ with a quality factor of $Q=10{,}000$ at room temperature ($T=300\unit{K}$). This results in a thermal noise limit of $F_{th}\approx 1 \cdot 10^{-14}\unit{N}$, which corresponds to the gravitational force exerted by a mass of the same size separated by $3\unit{mm}$ in distance. Obviously, such a simple estimate neglects the fact that the external gravitational force would have to be modulated in time, which in turn decreases the response of the sensor because of the finite modulation depth. As an order of magnitude estimate, however, it suggests that in principle it should be possible to exploit the sensitivity of state-of-the-art micro-mechanical devices to measure gravity between $\unit{mm}$-sized objects of $\unit{mg}$-scale mass, possibly even below that. Note that this is different from experiments that probe possible deviations from Newtonian gravity at short distances and that also involve small source masses \cite{Geraci2008Improved,Hamilton2015Constraints}. Their sensitivities and experimental configurations are targeted to put bounds on a modified force term, while our proposal is seeking to detect the (much weaker) signal generated by Newtonian gravity alone. 
In the following we introduce a concrete experimental design that is capable of doing exactly this. We first discuss the working principle, based on resonantly driving a micro-mechanical device by a time-dependent gravitational force that is created by a small, oscillating nearby source mass. We then analyze the technical requirements and the effect of other, non-Newtonian forces in the experiment. Finally, we provide an outlook on future possible applications of such an apparatus for improved precision measurements of Newton's constant and for a new generation of experiments at the interface between quantum physics and gravity, in which the quantum system itself can act as a gravitational source mass.

\section{Experimental scheme}
\begin{figure}[!htb]
	\centering
	\footnotesize
	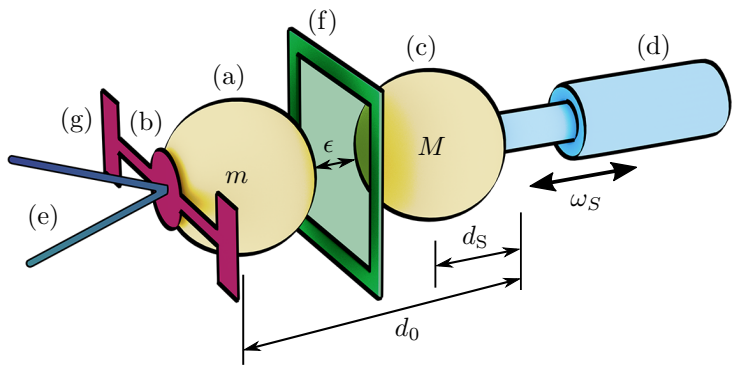
	\caption{Basic setup. A test-mass~$m$~(a) is loaded on a micromechanical device~(b). A source mass~$M$~(c) is located at a COM distance~$d_0$ from the test mass and is modulated through a drive motor~(d) with maximum amplitude~$d_\mathrm{S}$. The displacement of the test mass cantilever is read out optically (e). Other, non-gravitational forces are further suppressed by a shielding membrane~(f). (g) labels the mounting support structure.\label{fig:scheme}}
\end{figure}

Measuring the gravitational field of a massive object in an Earth-based laboratory has been implemented in various ways, ranging from torsional or linear test mass pendula over differential weight measurements (beam balance) to atom interferometry (see e.g. \cite{Speake2014TheSearch,Rothleitner2015Schwere} for recent reviews). Most of these methods rely on comparing different \emph{static} configurations of gravitational fields produced by a certain fixed geometry of source masses\footnote{A notable exception is the experiment by Gundlach et al. \cite{Gundlach2000MeasurementNewtons}, which uses a rotating source mass configuration and which also holds the current precision record for measuring the gravitational constant.}. In contrast to the $\unit{kg}$-size, macroscopic source masses of these experiments, we use $\unit{mg}$-scale, (sub-)millimeter sized objects. This allows us to easily create \emph{time-dependent} gravitational fields by actively shaking the source mass with respect to the test mass, which is attached to a micro-mechanical cantilever (\fref{fig:scheme}).  Modulating the distance between source mass and test mass at a frequency close to the mechanical resonance results in resonant driving of the test mass cantilever through a time-dependent gravitational field, i.e. in a gravitationally induced amplitude modulation of the test mass. Because of the resonant drive, the effect of the gravitational force is amplified by the mechanical quality factor~$Q$ (see \sref{sec:lin_harm_osc}). Other, unwanted forces can be neglected as long as the distance between the surfaces of the masses is kept sufficiently large, the background gas pressure is sufficiently low and a shielding membrane is placed in the gap between source and test mass (see \sref{sec:parameters}). Optical homodyning is employed to provide precision-readout of the (thermal-noise limited) position fluctuations of the test mass (see \sref{sec:readout}). We start our discussion by deriving the major signal contributions for this setup. 

\subsection{The linearized force-driven harmonic oscillator\label{sec:lin_harm_osc}}
Our system is composed of a spherical test mass loaded to a cantilever (\emph{test mass}) and a spherical driving mass (\emph{source mass}). In what follows we make the simplifying assumption that thermal noise contributions from internal friction of the cantilever \cite{Saulson1990ThermalNoise} can be neglected compared to velocity-dependent (viscous) damping, which is legitimate for a single-mode, narrowband detection scheme close to resonance. The dynamics are then governed by the equations of a one-dimensional harmonic oscillator with equilibrium position~$x^\prime$, internal damping rate~$\gamma^\prime = \omega^\prime_0/Q$, eigenfrequency~$\omega_0^\prime$ and mass $m$ that is driven in multiple ways: by motion (both deterministic and stochastic) of its support~$x_\mathrm{sup}$,  by various deterministic forces~$F_i$ that may depend on the total distance~$d_\mathrm{tot}$ and velocity~$\dot d_\mathrm{tot}$ between the center of mass (COM) of the oscillator and the driving system, and by mean-zero stochastic noise terms~$N_i$:
\begin{eqnarray}
\ddot x^\prime + \gamma^\prime \, \dot x^\prime + {\omega_0^\prime}^2 (x^\prime-x_\mathrm{sup}) = m^{-1} \left(\sum_i F_i(d_\mathrm{tot},\dot d_\mathrm{tot}) + \sum_i N_i \right) \mbox{.} \label{eq:original_oscillator}
\end{eqnarray}
As shown in \ref{app:lin_force_oscillator}, splitting the time-dependent distance $d_\mathrm{tot}$ into a static part $d_0$ and a mean-zero time-dependent part allows us to linearize the force terms around the non-deflected position of the test mass. Both the test mass position and frequency are shifted due to the presence of the force terms. Specifically, in case that Newtonian gravity ~$F_G=G  m  M  d_\mathrm{tot}^{-2}$ is the dominant force, the new effective frequency and position are\footnote{For simplicity we assume that the effective mass of the oscillator mode is identical to the gravitational mass.}  
\begin{eqnarray*}
\omega_0 = \left({\omega_0^\prime}^2 + 2 G M d_0^{-3} \right)^{1/2} \quad \mbox{and} \quad x = x^\prime - G M d_0^{-2} \omega_0^{-2} \mbox{.}
\end{eqnarray*}
These are typically the observables of torsional-balance experiments. For example, usual Cavendish-type experiments use centimeter-to decimeter-size source masses and a torsional pendulum operating at a resonance frequency of some milli-Hertz. With distances $d_0$ on the order of the size of the masses, say $d_0=10\unit{cm}$, $M\propto(d_0/2)^3$ and $\omega_0^\prime=10^{-3}\unit{Hz}$, one expects frequency shifts $\Delta\omega$ up to some hundreds of micro-Hertz and displacements up to millimeters, both of which can be reasonably resolved in precision measurements\footnote{The frequency shifts in actual measurements of $G$ are typically one order of magnitude higher, as the geometry of a torsion balance pendulum is only vaguely approximated by our $1$-dimensional, linear model.}. One way of reducing the mass further would be to reduce the experimental dimension $d_0$, which is however accompanied by an increase in resonance frequency $\omega_0^\prime$ (if we assume an unaltered spring constant). This results in a highly unfavourable scaling of the observable effects, since $\Delta\omega\propto 1/\omega_0^\prime$ and $\Delta x \propto d_0^{-2}\omega_0^{-2}$. In particular, using $d_0=1\unit{mm}$ and $\omega_0^\prime=10\unit{Hz}$ yields effective frequency and position shifts of tens of nanohertz and picometers, respectively, which is significantly more challenging to measure. For this reason, simply scaling down a Cavendish experiment is not sufficient to measure the gravitational effects of small source masses. 

Instead, we periodically modulate the gravitational potential created by a small source mass in order to resonantly enhance the amplitude response of a cantilever test mass. The power spectral density $S_{xx}$ of the (test mass) cantilever displacement is given by (\ref{app:lin_force_oscillator})

\begin{eqnarray}
\fl S_{xx}(\omega) = & & \left|\chi(\omega_\mathrm{S})\right|^2 \left|-2  G M/d_0^3 + {\omega_0^\prime}^2 \, T_\mathrm{S}(\omega_\mathrm{S})\right|^2 \frac{d_\mathrm{S}^2 \pi}{2} \left( \delta(\omega-\omega_\mathrm{S}) + \delta(\omega+\omega_\mathrm{S}) \right) \nonumber \\
\fl &+& \left|\chi(\omega)\right|^2 \left( {\omega_0^\prime}^4 T^2_\mathrm{E}(\omega)  S_{x_\mathrm{E} x_\mathrm{E}}(\omega)+ 2  \gamma  k_B T/m  \right) \nonumber \\
\fl &+& \mbox{further deterministic forces} + \mbox{further noise,} \label{eq:sxxmain}
\end{eqnarray}
where we define $\chi(\omega)=(\omega_0^2-\omega^2+\rmi \gamma \omega)^{-1}$ as the mechanical susceptibility. Here we take into account the Newtonian force~$F_G$ as well as thermal noise with power spectral density ${S_{NN}}_\mathrm{th} = 2  m  \gamma  k_B  T$. In addition, we assume a sinusoidal drive of the source mass with amplitude~$d_\mathrm{S}$. 

The first contribution is the gravitational effect in which we are interested. The second term is the mechanical drive from the deflection of the source mass transmitted through the supporting structure between source mass and test mass with transfer function $T_\mathrm{S}(\omega)$. The third contribution is due to the environmental vibrational noise~$S_{x_\mathrm{E} x_\mathrm{E}}(\omega)$ that is modified by a transfer function $T_\mathrm{E}(\omega)$. The last term describes Brownian motion of the test mass. 

In any actual experiment the measurement time $\tau$ is finite. In the following we make the experimentally justifiable assumption that the measurement bandwidth~$\Gamma=2
\pi /\tau$ is larger than the spectral width of the drive modulation and smaller than the mechanical width~$\gamma$, which simplifies the following analytical treatment of the expected measurement signal. The measured displacement power~$P_{xx}=\int_{\omega_\mathrm{S}-\Gamma/2}^{\omega_\mathrm{S}+\Gamma/2} S_{xx} \rmd \omega$ in the frequency band around $\omega_\mathrm{S}$ can be written as 
\begin{eqnarray*}
\fl P_{xx} = {P_{xx}}_G + {P_{xx}}_\mathrm{S} + {P_{xx}}_\mathrm{E} + {P_{xx}}_\mathrm{th} + \mbox{cross terms} + \mbox{further contributions}.
\end{eqnarray*}
For resonant driving ($\omega_\mathrm{S} = \omega_0$) and weak coupling ($\omega^\prime_0 \approx \omega_0$) one finds
\numparts \begin{eqnarray}
{P_{xx}}_G&= 2\pi \frac{Q^2}{\omega_0^4}  (G  M)^2  \frac{d_\mathrm{S}^2}{d_0^6} & \qquad \mbox{gravity,} \label{eq:p_contrib_g}\\ 
{P_{xx}}_\mathrm{th}&= 2 \frac{Q}{\omega_0^3} \frac{k_B T}{m} \Gamma & \qquad \mbox{thermal noise,} \label{eq:p_contrib_t}\\
{P_{xx}}_\mathrm{S}&= \frac{\pi}{2}  Q^2  T_\mathrm{S}^2(\omega_0)  d_\mathrm{S}^2 & \qquad \mbox{source mass drive,} \label{eq:p_contrib_d}\\
{P_{xx}}_\mathrm{E}&= Q^2  T_\mathrm{E}^2(\omega_0)  S_{x_\mathrm{E} x_\mathrm{E}}  \Gamma & \qquad \mbox{environmental vibrations,} \label{eq:p_contrib_e}
\end{eqnarray} \endnumparts
which are the relevant contributions to our expected signal.

\subsection{Signal strength, thermal noise and force contributions\label{sec:signalnoiseforces}}
When neglecting all other noise sources, thermal noise becomes a fundamental hurdle for seeing the effect of gravitation on a small scale. A few interesting insights can be obtained by comparing the scaling of the gravitational contribution~\eref{eq:p_contrib_g} with the one of thermal noise~\eref{eq:p_contrib_t}. First, while the thermal noise decreases for larger test masses~$m$, the gravitational contribution does not depend on the size of the test mass but only of the source mass~$M$ (as is expected from the weak equivalence principle). Therefore, to ensure that gravity dominates over the thermal noise, either source mass or test mass (or both) can be increased. In our case we want to keep the source mass small and hence can increase the test mass. One trade-off that needs to be considered in this case is the strong scaling of the gravitational contribution with the COM distance between test and source mass ($d_0^{-6}$), which is likely to increase when increasing the size of the test mass. Second, the thermal noise scales linearly with the bandwidth, whereas the gravitational contribution does not - in other words, a longer measurement time will decrease the stochastic thermal noise when compared to the steady-state oscillatory signal. Third, because of the explicit $Q$-dependence of the Brownian force noise (due to the fluctuation-dissipation theorem), both contributions scale differently with mechanical quality~$Q$: quadratic for the gravitation and linear for the thermal noise part. Hence, increasing $Q$ will not only lift the absolute thermal noise level but will also improve the signal to noise ratio between gravitational signal and thermal noise. For the same reason, scaling is also different in mechanical frequency, specifically $\omega_0^{-4}$ for the gravitational and $\omega_0^{-3}$ for the thermal noise part. Operating at lower mechanical resonance frequencies is therefore favorable. In summary, our gravitational sensing approach should allow for achieving a good signal to noise ratio. Although the relative noise contribution increases with smaller (test) masses and larger COM distances $d_0$, this can be compensated for by larger mechanical quality factors $Q$ and longer measurement times.

\subsection{Parameters\label{sec:parameters}}
\begin{figure}[!htb]
	\centering
	\includegraphics{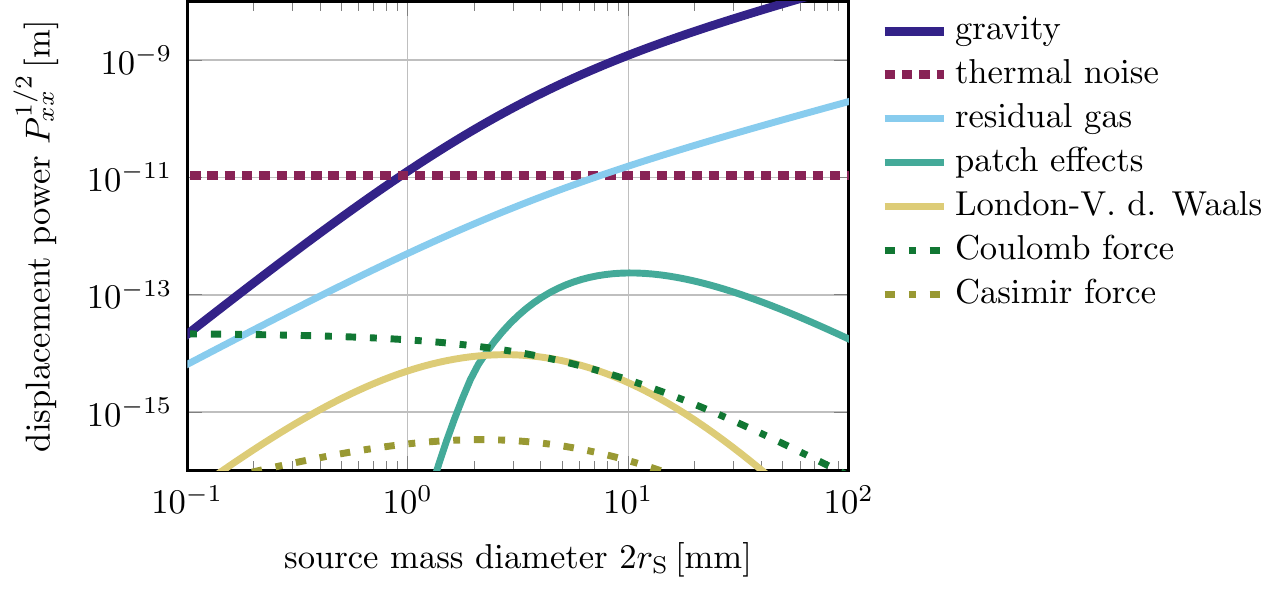}
	\caption{Signal contribution of gravity, thermal noise and various forces as a function of source mass diameter at $T=300\unit{K}$ and $\Gamma=2 \pi / 3600\unit{s}$. For the Coulomb force we assumed 200 surface charges per mass with opposing charges located at the closest position. The London-Van der Waals force is shown for the worst possible material properties. Residual gas is shown for a pressure of $10^{-8}\unit{hPa}$. The plot assumes a minimal surface distance of $\epsilon=0.5\unit{mm}$ and an optimal modulation amplitude (\sref{sec:drivemass}). The expressions for the additional forces shown can be found in \ref{app:nongravforces}.  
	\label{fig:plot_forces}}
\end{figure}

With the expressions derived in \sref{sec:lin_harm_osc}, we now assess the feasibility of the experiment for a realistic parameter regime. \Fref{fig:plot_forces} shows the signal contribution of thermal noise and gravity as a function of the source mass radius. Here we assume a test mass of the same size as the source mass, a test mass cantilever of frequency $\omega_0=50\unit{Hz}$ and mechanical quality factor $Q=2 \cdot 10^4$, and an integration time of one hour, i.e. ~$\tau=2\pi/\Gamma=3600\unit{s}$. The material of choice is gold due to its high density ($\rho_{Au}=1.93\cdot 10^4\unit{kg/m^3}$), purity and homogeneity \cite{Montgomery2011SRM}. In addition, we assume that the minimal distance between the surfaces of source and test mass is $\epsilon=0.5\unit{mm}$ and we choose an optimal drive amplitude for the source mass modulation (see \sref{sec:drivemass}). For these (conservative) settings, a signal to noise ratio of 1 is reached for a source mass radius of $500\unit{\mu m}$, which in case of gold corresponds to a source mass weight of about $10\unit{mg}$. For our further considerations we want to leave some overhead for unaccounted experimental noise sources and hence choose a source mass radius of $1\unit{mm}$, where the gravitational contribution is about 6 times higher than the thermal noise. A gold sphere of this size has a volume of $4.2 \unit{{mm}^3}$ and a mass of $80.9\unit{mg}$, which is still three orders of magnitude smaller than the smallest reported attractor masses in a laboratory based experiment \cite{Ritter1990Equivalence,Gillies2015AttractingMasses}. \Fref{fig:plot_forces} also shows the contribution of other residual forces. For unwanted Coulomb forces we assume 200 surface charges per mass with opposing charges located at the closest position on each sphere\footnote{Note that for the parameters discussed here $5{,}000$ charges of that type would be required for generating a Coulomb force that equals the thermal noise contribution.}. The London-Van der Waals force contribution that is shown is estimated for the worst possible material properties and the effects of residual gas scattering is shown for a pressure of $10^{-8}\unit{mbar}$. Details of the calculations are shown in \ref{app:nongravforces}. Another possible effect is non-contact friction due to time-dependent electric fields (\emph{patch potentials}), which is a known effect for conducting surfaces \cite{Burnham2016WorkFunction,Stipe2001Noncontact}. Due to the relatively large distance $\epsilon$ between the test and source mass surfaces, such fields can be shielded by a membrane between the two masses \cite{Weld2007Design}.

\section{Technical requirements\label{sec:tecreq}}
In our previous analysis we have assumed that the test mass cantilever fluctuations are dominated by thermal noise, i.e. all other noise contributions need to be sufficiently small. We identify and discuss four main technical challenges to achieve this requirement:
\begin{enumerate}
\item fabrication of a test mass cantilever that meets the mechanical criteria above,
\item sufficient vibration isolation against environmental and drive noise,
\item low-noise readout of the test mass cantilever motion, and
\item a mechanical source mass drive that does not introduce significant additional vibrational noise.
\end{enumerate}

\subsection{Test mass cantilever}
The last two decades have seen dramatic improvements in the fabrication and performance of nano- and micro-mechanical devices \cite{Zwickl2008SiNMembranes,Imboden2014Dissipation}. For our experiment, we are considering a micromechanical system that is mass-loaded with a $1\unit{mm}$-radius gold sphere, thereby forming a test mass cantilever at the target frequency of $\omega_0\approx 50 - 100\unit{Hz}$ (see \ref{app:fem}). An outstanding question is the achievable mechanical quality for such a structure. In the context of atomic force microscopy (AFM) with colloidal probes, polystyrene  microbeads of glass, polystyrene, polyethylene and other materials have been successfully attached to cantilevers while maintaining typical AFM cantilever quality factors in the ten-thousands to millions \cite{Butt2005AFM}. Our experiment deals with significantly more massive objects, which will require a relatively large attachment area. As we could not find consistent values for the bulk quality factors of the high-density metals gold and lead, a rough estimate was gathered in a piezomechanical S21 gain/loss measurement (\ref{app:fem}). With a measured value of $Q\approx450$ for gold we assumed $100$ as a worst case estimate. With such low mechanial quality it is important to avoid deformation of the test mass as a mode shape contribution of the relevant COM mode. This requires a careful design of the cantilever geometry. Finite element modeling (FEM) methods provide the means to analyze mode shapes and from that estimate effective $Q$-values of compound cantilever systems, which can be used to optimize geometries with regard to test mass deformation. It turns out that in a simple cantilever-geometry the mechanical quality of the material directly at the bonding surface between cantilever and test mass can have a huge negative impact on the overall quality of the compound system. This can be circumvented by changing the cantilever geometry such that deformation of the bonding surface is avoided, or by attaching the test mass to the cantilever using an adhesion layer with low internal losses \cite{Schediwy2005HighQBonding}. A specific example is presented in \ref{app:fem}. Assuming $Q$s of $30{,}000$ for an AlGaAs cantilever \cite{Cole2012CrystallineMirrors}, $300$ for the adhesive \cite{Schediwy2005HighQBonding} and $100$ for the test mass one obtains the overall dissipation by adding up the loss angles and at the same time scaling their contribution with the mode stress derived from FEM simulations, which yields quality factors of the mass-loaded structure of at least $Q \approx 24{,}000$. A brief estimation of damping through Brownian noise from gas impacts is presented in \ref{app:gasimp}.

\subsection{Seismic isolation}
The test mass cantilever displacement is subject to additional external noise sources, in particular seismic noise of the environment, $S_{x_\mathrm{E} x_\mathrm{E}}$, and mechanical backaction of the source mass displacement, which is coupled through the mechanical support structure via the transfer functions $T_\mathrm{E}$ and $T_\mathrm{S}$, respectively (see \eref{eq:sxxmain}). Their contributions can therefore be damped by additional vibration isolation of the test mass cantilever. To achieve a suppression well below the thermal noise limit in the measured signal power requires $P_{xx,T} > P_{xx,E}, P_{xx,D}$ (see \eref{eq:p_contrib_t}, \eref{eq:p_contrib_d}, \eref{eq:p_contrib_e}), yielding
\begin{eqnarray*}
T_\mathrm{E}(\omega)  < \left(\frac{2 k_B T}{Q m \omega_0^3} \right)^{1/2}   S_{x_\mathrm{E} x_\mathrm{E}}^{-1/2} \approx 10^{-7} \mbox{,}
\end{eqnarray*}
where we assume typical laboratory noise of $S_{x_\mathrm{E} x_\mathrm{E}}^{1/2}\approx 10^{-8}~\unit{m/Hz^{1/2}}$ at $50\unit{Hz}$, and
\begin{eqnarray*}
 T_\mathrm{S}(\omega_\mathrm{S}) < 2 \left( \frac{k_B T  \Gamma }{\pi Q m \omega_0^3}  \right)^{1/2}  d_\mathrm{S}^{-1}\approx 10^{-18}.
\end{eqnarray*}
The first term requires an isolation of the test mass cantilever platform from seismic noise by at least $70\unit{dB}$ at around $50-100 \unit{Hz}$, which is clearly within current state of the art. For example, a combination of multiple passive and actively controlled suspension stages in gravitational wave detectors routinely achieve seismic isolations of $100\unit{dB}$ and better at even lower frequencies. For our case, already a dual-stage passive spring-pendulum system should be sufficient to achieve the required levels of isolation at $50 \unit{Hz}$ \cite{Matichard2015SeismicIsolation}. 

The second term seems to impose a significant challenge, but one should bear in mind that we consider here the contribution of the source mass displacement that is due to mechanical backaction on the support structure of the experiment. There are several strategies to minimize this. First, mechanical unbalance can be compensated for by having a second mass counter-moving against the first (\sref{sec:drivemass}). Second, the source mass drive platform can be physically separated from the test mass cantilever platform. This requires a separate vibration isolation (spring pendulum) system, which couples to the test mass cantilever platform only via the large mass of the vacuum tank that hosts the experiment. Finally, one can even envision a complete mechanical isolation of the source mass by levitating and driving it in external fields \cite{Goodkind1999Superconducting}.

\subsection{Optical readout\label{sec:readout}}
For our envisioned parameter regime, gravitational driving will result in signal noise powers of the test mass displacement of $S_{xx}^{1/2}(\omega_0)\approx10^{-10}m/Hz^{1/2}$ on resonance and $S_{xx}^{1/2}(\omega=0)\approx10^{-14}m/Hz^{1/2}$ off resonance. Optical interferometry is a convenient and well-established way to read out such small signal levels. In essence, the displacement $\delta x$ is converted into an optical phase modulation $\delta\phi=2\pi \delta x /\lambda$, which can be measured either directly as amplitude modulation in a balanced interferometer or via optical homodyne detection \cite{Bachor2004QuantumOptics}. The challenge for our experiment is to obtain this sensitivity at small frequencies in the audio band. One has to consider the following noise sources: classical amplitude- and phase noise of the laser source, quantum noise (shot noise and backaction), electronic noise from the detection circuit, and residual amplitude modulation in the readout architecture.

\begin{figure}[!htb]
	\centering
	\includegraphics{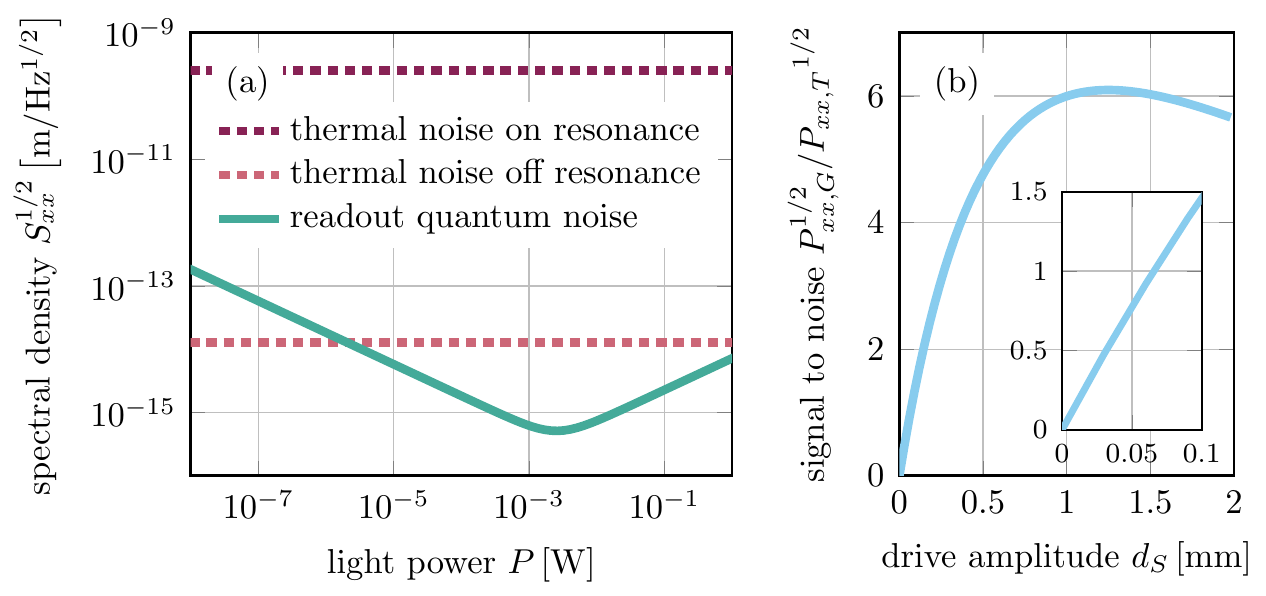}
	\caption{(a) Readout quantum noise (dominated by shot noise for lower powers and backaction noise for higher powers), thermal noise and on-resonance ($\omega=\omega_0$) and off-resonance ($\omega=0$). (b) Relative signal for a drive amplitude deviating from the optimum ${d_\mathrm{S}}_\mathrm{opt}=1.25\unit{mm}$. The inset shows the behavior for small amplitude drives up to $100 \unit{\mu m}$.\label{fig:plot_power_drive}}
\end{figure}

The optical source noise is composed of amplitude noise, which we can circumvent by optical homodyning and measuring the phase qudrature, and phase noise, which can essentially be converted to amplitude noise using polarization optics. Quantum noise in an optical position measurement is due to intrinsic photon number fluctuations in a laser beam (\emph{shot noise}), which contributes statistical noise both in photon counting and in actual displacement due to radiation-pressure induced momentum transfer (\emph{backaction}). This results in the well-known standard quantum limit for continuous position sensing, which resembles the working point of maximal achievable sensitivity in the presence of quantum noise \cite{Caves1980Quantum,Braginsky1995QuantumMeasurement,Clerk2010QuantumNoise}. For reasons of practicality, we will discuss a readout scheme that does not invoke a cavity configuration but consists only of a two-path interferometer operated at an optical input power $P$. In this case, the added shot-noise contribution to the displacement measurement is \cite{Clerk2010QuantumNoise} ${S_{xx}}_\mathrm{shot}=\hbar \lambda c(32 \pi P)^{-1}$, which is an effect of photon counting and hence does not depend on the mechanical susceptibility of the test mass. In contrast, the backaction contribution to the displacement noise is ${S_{xx}}_\mathrm{back}=\left|\chi(\omega_\mathrm{S})/2\right|^2 (\hbar/m)^{2} {S_{xx}^{-1}}_\mathrm{shot}$, which is amplified by the mechanical response. \Fref{fig:plot_power_drive}(a) compares the noise contributions of photon shot noise and backaction (at $\omega=\omega_0$) to the thermal noise floor on and off the mechanical resonance. Because of the large mass of the test mass system backaction noise can be neglected for all reasonable parameter regimes. 

For resonant detection it is sufficient to suppress the frequency-independent shot noise well below the thermal noise contribution at the mechanical frequency, i.e. ${S_{xx}}_\mathrm{thermal}(\omega_0)\gg{S_{xx}}_\mathrm{shot}$, yielding
\begin{eqnarray*}
P   \gg  \frac{\hbar\lambda c}{64 \pi} \frac{m\omega_0^3}{k_B T}\frac{1}{Q} \approx 5\unit{fW}
\end{eqnarray*}
It may also be useful to fully resolve the thermal noise of the oscillator, allowing for example for off-resonant detection schemes. This requires suppression of the shot noise contribution well below the off-resonance thermal noise, resulting in  
\begin{eqnarray*}
P   \gg  \frac{\hbar\lambda c}{64 \pi} \frac{m\omega_0^3}{k_B T}Q \approx 2\unit{\mu W},
\end{eqnarray*}
which was derived using the thermal noise contribution from \eref{eq:sxxmain} at $\omega=0$. 

Although detection in the audio-band is challenging because of unavoidable low-frequency fluctuations in the setup \cite{McKenzie2002Squeezing,Vahlbruch2008Squeezed,Stefszky2012Generation}  compact interferometric readout schemes have been demonstrated that operate with the desired sensitivity \cite{Rugar1989FiberOptic,Smith2009FiberOptic,Paolino2013Quadrature}. For example, by combining a robust homodyning architecture with a large-bandwidth probe laser a recent experiment has reported a displacement sensitivity of $4\cdot10^{-14}\unit{m/Hz^{1/2}}$ above $20 \unit{Hz}$ at laser powers of $10\unit{\mu W}$. \cite{Smith2009FiberOptic}

\subsection{Source mass\label{sec:drivemass}}
We set the \emph{minimal distance}~$\epsilon$ between the test mass and source mass surfaces to a fixed value. One can then derive an optimal distance $d_0$ and driving amplitude $d_s$. By using $d_0=d_\mathrm{S}+r_\mathrm{T}+r_\mathrm{S}+\epsilon$ ($r_\mathrm{T/S}$: test/source mass radius) and the fact that the Newtonian contribution to the measurement signal scales with $d_\mathrm{S}^2\,d_0^{-6}$, we find the optimum to be
\begin{eqnarray*}
{d_0}_\mathrm{opt} = \frac{3}{2} \, \left(r_\mathrm{T} + r_\mathrm{S} + \epsilon \right), \quad {d_\mathrm{S}}_\mathrm{opt} = \frac{1}{2}\,\left(r_\mathrm{T} + r_\mathrm{S} + \epsilon\right)
\end{eqnarray*}
For the parameters discussed above ($\epsilon=0.5\unit{mm}, r_\mathrm{T}=r_\mathrm{S}=1\unit{mm})$  the optimal values are ${d_\mathrm{0}}_\mathrm{opt}=3.75\unit{mm}$ for the COM distance and ${d_\mathrm{S}}_\mathrm{opt}=1.25\unit{mm}$ for the actuation amplitude. Achieving smooth driving of the source mass at around $50\unit{Hz}$ at such amplitude, with sufficiently large lifetime ($10^8$ to $10^9$ cycles) and without adding significant stray fields represents a substantial engineering challenge. State-of-the art piezoelectric actuators fall short of the required drive amplitude by at least one order of magnitude and available actuated positioning platforms do not achieve the desired accelerations between $123\unit{ms^{-2}}$ ($50\unit{Hz}$) and $493\unit{ms^{-2}}$ ($100\unit{Hz}$). One possible drive mechanism could be a spring-mounted electromagnet, which in principle allows actuation at small input power and therefore small stray fields and heating \cite{Drummer2008Magnetventil}. Alternatively, one can operate at a smaller drive amplitude. \Fref{fig:plot_power_drive}(b) shows the resulting decrease in signal strength. With the current parameter settings, an overall signal to noise ratio of 1 can be achieved with a drive amplitude of around $70\unit{\mu m}$.

In order to position the source mass, an optical scheme can  be employed to read out the relative distance between the test mass and the source mass in all spatial directions. This can be accomplished, for example, with quadrant diodes at the source mass stage paired with focused light beams from the test mass stage. If higher precision is required, one might think of interferometric schemes or even techniques involving optical cavities, which would in principle allow to push relative positioning errors into the sub-nanometer regime. The displacement of the source mass could be read out in a similar fashion, however the technical implementation will depend on the mechanism used for the source mass drive.

\section{Further developments}
With the choice of realistic parameters given above our micromechanical method should allow to demonstrate gravitational coupling between masses below 100 mg, i.e. three orders of magnitude below the current smallest source mass values. We envision several strategies how the estimated sensitivity can be improved even further. One possibility is to replace the spherical test and source masses by objects whose shapes are optimized for the task of detecting a modulated $r^{-2}$ force at a given minimal distance. Rough numerical estimations indicate that this could yield a gain of up to one order of magnitude in signal power. Another possibility is to increase the mechanical quality factor~$Q$ of the test mass cantilever, which scales linearly with the signal to noise (power) ratio. At this stage most of our assumptions on the cantilever performance have been rather conservative and the actual performance might turn out to be much higher than $Q=2 \cdot 10^4$. Ultimately, both mass suspensions could be replaced by magnetic traps in order to achieve levitation of source and test mass. Such levitated systems offer significantly higher $Q$ values due to their strongly suppressed environmental coupling \cite{Romero2012LevitatingSupercond,Cirio2012Magnetomechanics}.
These improvements could open up interesting application areas for our scheme. On the one hand, high-precision measurements of the gravitational field of small source masses offer a completely different approach to determine Newton's constant, possibly less sensitive to systematic errors present in experiments with macroscopic source masses. On the other hand, combining the sensitivity to gravitational coupling between microsopic source masses with the ever growing ability to achieve quantum control over their center of mass motion will lead to a completely new generation of experiments at the interface between quantum physics and gravity.

\subsection{Measurement of the gravitational constant}
The accurate determination of Newton's constant~$G$ has become a highly debated subject \cite{Reich2010GWhizzes,Quinn2016Balance}. Although some experiments are now reaching precision levels up to $\Delta G/G\approx 1 \cdot 10^{-5}$ \cite{Gundlach2000MeasurementNewtons}, different implementations continue to disagree in the absolute value of G by multiple standard deviations \cite{Mohr2012Codata,Faller2014ErrorBudget,Rothleitner2015Schwere}. A significant, if not dominant, contribution to the error budget of most of these measurements is due to uncertainties associated with the manufacture of the macroscopic source masses and their incorporation and use in the experimental arrangements. This includes suspension noise \cite{Saulson1990ThermalNoise} as well as inaccuracies in the center of mass distance between test and source mass due to, for example, inhomogeneities or temperature fluctuations. In addition, long integration times require a detailed understanding of all long-term systematics in these experiments. Alternative approaches for measuring $G$ may therefore provide helpful insights. One is cold-atom interferometry \cite{Fixler2007AtomInterferometer,Rosi2015Precision}, where a precision of $\Delta G/G\approx 1 \cdot 10^{-4}$ has recently been demonstrated that was mainly limited by the position measurement of the atoms with respect to a macroscopic tungsten source mass \cite{Rosi2015Precision}.

Our approach involves a centimeter-scale experimental architecture, a microscopic source mass and short integration times of only hours. This combination reduces conventional sources of errors in precision measurements of $G$, since the small volumes enable a better control of positioning and density inhomogeneities of the masses as well as of temperature fluctuations. In addition, short integration times may allow for a systematic study of the influence of fluctuations of other spurious external forces that give rise to systematic errors in long-term experiments. A remaining challenge is to achieve a measurement precision that is competitive with experiments involving macroscopic source masses. One straight-forward way to achieve this is to increase the size of the masses, which will both boost the gravitational signal and decrease the thermal cantilever noise. For example, following our analysis above and using 10mm radius spheres instead of 1mm with otherwise unaltered parameters would result in a signal to noise ratio beyond $10^6$, i.e. a precision of $\Delta G/G < 1 \cdot 10^{-6}$. Ultimately, operating the experiment in a low-temperature environment would in principle allow for even higher precision, provided that all technological challenges of low-frequency, cryogenic vibration isolation can be addressed in future experiments. Some third-generation gravitational wave detector designs have already been studying cryogenic scenarios \cite{Somiya2012DetectorConfiguration,Punturo2010EinsteinTelescope}.

\subsection{How does a quantum system gravitate?}
Although the predictions of both quantum theory and general relativity are extremely well confirmed by experiment, interfacing these two theories belongs to one of the outstanding big challenges of modern science. Notwithstanding the conceptual and mathematical hurdles in writing down a full quantum theory of gravity, the number of available experiments that probe the interface between quantum physics and gravity is also extremely sparse. One type of experiments focus on observations over astronomical distances, which may reveal imprints of quantum gravity effects \cite{Amelino2013QuantumSpacetime,Kiefer2012QuantumGravitational}. The other type of experiments exploit the availability of continuously improving high-precision lab-scale experiments \cite{Lammerzahl2006TheSearchA,Lammerzahl2006TheSearchB}. The latter ones fall essentially into two categories: they are either genuine quantum tests in the limit, where Newtonian gravity acts as a constant classical background field \cite{Colella1975QuantumInterference,Kasevich1991Atomic,Nesvizhevsky2002QuantumStates,Jenke2011Spectroscopy,Biedermann2015TestingGravity}, or they are tests of genuine gravity effects measured through high-precision quantum experiments \cite{Pound1959RedShift,Blatt2008NewLimits,Chou2010OpticalClocks}. In other words, thus far all of these laboratory scale experiments have been using quantum systems as test masses in external gravitational fields. Using quantum systems as gravitational source masses would establish a qualitatively new type of experiment. Obviously, this will require quantum control over the motion of sufficiently massive objects and at the same time the experimental sensitivity to their gravitational forces. In this context, our micromechanics platform presented in this paper can be seen as a top-down approach for designing such future experiments. The lowest masses and shortest timescales above which gravitational coupling can be observed will be an important benchmark for both mass and coherence time of future quantum experiments. In the most optimistic scenario, the combination of force sensitivity and coherence time will eventually enable the quantum regime of gravitational source masses, for example by demonstrating gravitationally induced entanglement as suggested by Feynman \cite[p.~250]{DeWitt2011TheRole}.

\section{Summary}
We have introduced a micromechanical method to measure gravitational coupling between small masses. Current state of the art technology should allow for a proof-of-concept demonstration for objects on the  scale of millimeters and tens of milligrams, which already improves the current limit for sensing the gravitational field of a small source mass by three orders of magnitude. With further improvements this method provides an alternative high-precision measurement of the gravitational constant, which may be less subject to conventional source-mass related disturbances of other approaches. Finally, in the long run, the ability to extend the control over gravitational coupling into the microscopic domain may enable a new generation of quantum experiments, in which the source mass character of the quantum systems start to play a role.

\ack

We would like to thank Rana Adhikari, Garrett Cole, George Gillies, Sebastian Hofer, Harald L\"uck, Conor Malcolm Mow-Lowry, Ralf Riedinger, and Tobias Westphal for insightful discussions and advice, and Stephan Puchegger for performing initial $Q$ measurements on the adhesives. We acknowledge support by the European Commission (cQOM), the European Research Council (ERC CoG QLev4G), and the Austrian Science Fund (FWF) under project F40 (SFB FOQUS). J. S. is supported by the FWF under project W1210 (CoQuS).

\newpage

\appendix

\section{Derivation of power contributions\label{app:lin_force_oscillator}}
Starting from \eref{eq:original_oscillator}, we split the time-dependent distance into $d_\mathrm{tot} = d_0 + x^\prime - x_\mathrm{S}$, where $d_0$ is the (static) COM distance with both masses being non-deflected and $x_\mathrm{S}$ is the relative drive motion. We can then approximate the force terms as
\begin{eqnarray}
\sum_i F_i &\approx \underbrace{\sum_i \left. F_i\right|_{d_0}}_{ \equiv m \, \varsigma} + \underbrace{\sum_i \left. \partial_{x^\prime-x_\mathrm{S}} F_i \right|_{d_0}}_{\equiv m \, \xi} \, (x^\prime-  x_\mathrm{S}) + \underbrace{ \sum_i\left. \partial_{\dot x^\prime-\dot x_\mathrm{S}}F_i\right|_{d_0}}_{\equiv m\,\zeta} \, (\dot x^\prime- \dot x_\mathrm{S}) \nonumber
\end{eqnarray}
with $\left. \right|_{d_0}$ meaning \emph{evaluated at $d_\mathrm{tot}{=}d_0$} and with $\varsigma$, $\xi$ and $\zeta$ being defined as the relevant amplitudes for convenience. Here we consider the series expansion to first order. Note, however, that taking into account higher orders of the source mass deflection is possible without changing the mathematical nature of the problem. Plugging this into \eref{eq:original_oscillator} yields
\begin{eqnarray}
\ddot x + \gamma \, \dot x + \omega_0^2 \, x = {\omega_0^\prime}^2 \, x_\mathrm{sup} - \xi \, x_\mathrm{S} - \zeta \,  \dot x_\mathrm{S} + \sum_i N_i/m \label{eq:oscillator_two}
\end{eqnarray} 
where we defined $\omega_0^2 {=} {\omega_0^\prime}^2 {-} \xi$, $\gamma {=} \gamma^\prime {-} \zeta$ and $x{=}x^\prime{-}\varsigma \, \omega_0^{-2}$ as the new effective position, damping and equilibrium position due to the presence of deterministic forces.

Converting \eref{eq:oscillator_two} into Fourier space yields 
\begin{eqnarray}
\tilde x =  -  \chi \, A\, \tilde x_\mathrm{S} + \chi \, \left( \sum_i \tilde N_i/m + {\omega_0^\prime}^2 T_\mathrm{E} \, \tilde x_\mathrm{E} \right). \label{eq:oscillator_three}
\end{eqnarray} 
Here, $\chi(\omega)=(\omega_0^2-\omega^2+\rmi  \gamma  \omega)^{-1}$ is the \emph{susceptibility} and $A(\omega)=\xi + \rmi  \omega \zeta + {\omega_0^\prime}^2 \, T_\mathrm{S}(\omega)$ is the amplitude of the system. We split up the support motion into an environmental statistical noise and a drive contribution with their respective transfer functions, $\tilde x_\mathrm{sup} = T_\mathrm{E}(\omega) \, \tilde x_\mathrm{E} + T_\mathrm{S}(\omega) \, \tilde x_\mathrm{S}$ where $T_\mathrm{E}$ and $T_\mathrm{S}$ are the frequency-dependent real functions that describe how a finite amplitude excitation is modulated after progressing from the point of deflection to the test mass oscillator. The first term of \eref{eq:oscillator_three} represents the deterministic contributions and the second term are the statistical noise contributions. 

In order to relate the Fourier transform to the accessible quantities in a measurement and to be able to compare noise terms and deterministic contributions, it is useful to consider the common definition of the \emph{power spectral density} \cite{Clerk2010QuantumNoise} of a physical quantity $x$,
\begin{eqnarray}
S_{xx} \equiv \lim_{T \to \infty} \left< \left| _T \tilde x(\omega) \right|^2\right>, \quad \mbox{with} \quad  _T \tilde x(\omega) \equiv (2T)^{-1/2} \int_{-T}^{+T} x\,\rme^{\rmi \, \omega \, t} \, \rmd t \label{eq:sd_one}.
\end{eqnarray}
By rewriting the \emph{windowed Fourier transform}~$ _T \tilde x$ as
\begin{eqnarray}
\eqalign{ _T \tilde x(\omega) = (2\pi)^{-1}\,(2T)^{-1/2}  \,  _T\tilde h(\omega) \ast \tilde x(\omega) \cr
 \mbox{with} \quad _T h(t) \equiv \cases{%
1&for $t \in [-T;T]$ \\%
0&else.\\} }
\label{eq:windowedfourier}
\end{eqnarray}
we may determine the windowed Fourier transform from the infinite Fourier transform. We now plug \eref{eq:windowedfourier} and \eref{eq:oscillator_three}  into  \eref{eq:sd_one} and assume that all sources of force noise~$N_i$ as well as the environmental noise $x_\mathrm{E}$ are uncorrelated. With a sinusoidal drive of the form 
\begin{eqnarray*}
x_\mathrm{S}(t) \equiv d_\mathrm{S} \cos(\omega_\mathrm{S} \, t)
\end{eqnarray*}
with drive frequency~$\omega_\mathrm{S}$ we can calculate the full spectrum as
\begin{eqnarray}
\eqalign{S_{xx} = \left|\chi(\omega_\mathrm{S})\right|^2 \, \left|A(\omega_\mathrm{S})\right|^2 \frac{d_\mathrm{S}^2 \pi}{2} \left( \delta(\omega-\omega_\mathrm{S}) + \delta(\omega+\omega_\mathrm{S}) \right) \\
+ \left|\chi(\omega)\right|^2 \left( {\omega_0^\prime}^4 \tilde T^2_\mathrm{T}(\omega) \, S_{x_\mathrm{E} x_\mathrm{E}}(\omega)+ \sum_i {S_{NN}}_i(\omega)/m^2  \right) } \label{eq:sd_two}
\end{eqnarray}
where the first line represents the deterministic contributions and the second line the noise terms.

There are two things to notice in \eref{eq:sd_two}: First, with $A(\omega_\mathrm{S})$ entering quadratically, there are not only quadratic amplitudes of all deterministic force terms, but also cross-terms of the various forces, e.g. a cross-term between gravity and the Coulomb force. As all forces drive the test mass with the same frequency~$\omega_\mathrm{S}$, they will be impossible to distinguish. Therefore it will be necessary to properly shield them from the test mass. Second, every force or noise is modified by the mechanical susceptibility when acting on the test mass. This means that improving the mechanical quality factor of the test mass might not necessarily enable a measurement of gravity if other effects dominate over the gravitational contribution. 

When taking into account the finite bandwidth~$\Gamma=2 \pi / \tau$ of an actual measurement of total time $\tau$, we can further process the result. Assuming that $\Gamma$ is larger than the spectral width of the drive modulation and smaller than the mechanical width~$\gamma$ simplifies the analytical treatment of the above expression. Then we can write the measured displacement power~$P_{xx}$ in the frequency band around $\omega_\mathrm{S}$ as
\begin{eqnarray}
\fl P_{xx} & \equiv & \int_{\omega_\mathrm{S}-\Gamma/2}^{\omega_\mathrm{S}+\Gamma/2} S_{xx} \rmd \omega \nonumber \\
\fl & \approx & \left|\chi(\omega_\mathrm{S})\right|^2  \, \left( 
\left|A(\omega_\mathrm{S})\right|^2 \frac{d_\mathrm{S}^2 \pi}{2} 
+ {\omega_0^\prime}^4 T_\mathrm{E}^2(\omega_\mathrm{S}) \, S_{x_\mathrm{E} x_\mathrm{E}}(\omega_\mathrm{S}) \, \Gamma  + \sum_i {S_{NN}}_i(\omega_\mathrm{S}) \, m^{-2} \, \Gamma  \right)   \label{eq:sd_three}
\end{eqnarray}

\section{Explicit form of non-gravitational force contributions\label{app:nongravforces}}
For the simple case of two spherical masses we list the expected force contributions. The definitions of $d_0$ as the equilibrium center of mass distance and $r_\mathrm{T}$, $r_\mathrm{S}$ as the test and source mass radii are common among all terms.

For the Coulomb force~$F_e$ we consider the (worst) case in which all relevant charges are located at the closest possible positions on the sphere surfaces:
\begin{eqnarray*}
F_e = \frac{1}{4 \pi \epsilon_0} \frac{q_1 q_2}{(d_\mathrm{tot} - r_\mathrm{T} - r_\mathrm{S})^2}
\end{eqnarray*}
with the attracting charges $q_1$ and $q_2$ on the test- and source mass. Charge accumulation on suspended test masses has been studied extensively in the context of gravitational wave detectors \cite{Hewitsen2007ChargeMeasurement,Mitrofanov2002Variation}. For the case of large ($\unit{cm}$-scale) fused silica mirrors surface charge densities up to $10^6\unit{ C/cm^{2}}$ have been observed directly after evacuation, probably due to friction-related effects during the pumping process. For our geometry (spherical masses of $2\unit{mm}$ diameter) this would result in  approx. $30{,}000$ charges per mass. However, static charging of this type can be removed through various ways, either by discharging through electrical contact or by UV light illumination \cite{Shaul2008ChargeManagement,Hewitsen2007ChargeMeasurement}. Further potential charging mechanisms may arise from cosmic radiation \cite{Braginsky2006NoiseCosmic}. Following \cite{Buchman1995ChargeMeasurement} one can use the Bethe-Bloch formula to calculate the energy range of protons and electrons that would, after penetrating the laboratory walls and the vacuum tank, come to stop in the test mass and potentially charge it. One can compare this to the tabulated particle background at sea level \cite[ch.~28]{Olive2014ReviewOfParticlePhysics}, which sums up to between $0.1$ and $0.01 \unit{m^{-2}s^{-1}}sr^{-1}$ scattering events per second at the relevant energies well below $1 \unit{GeV}$ (i.e. close to the material critical energy). Again, for our geometry this results in approx. $10^{-5}\unit{s^{-1}}$ scattering events with each mass or on the order of one ionizing scattering event per five to fifty hours. This estimate is consistent with long-term measurements on silica test masses in high vacuum, which report a monotonic charging rate of up to $10^5$ electrons per $cm^2$ per month \cite{Mitrofanov2002Variation}.

For the London-Van~der~Waals force the following expression holds \cite{Hamaker1937LondonVanDerWaals}:
\begin{eqnarray*}
\fl F_\mathrm{VDW} = \frac{32}{3} A \frac{r_\mathrm{T}^3 r_\mathrm{S}^3 d_\mathrm{tot}}{(d_\mathrm{tot}-r_\mathrm{T}-r_\mathrm{S})^2(d_\mathrm{tot}-r_\mathrm{T}+r_\mathrm{S})^2(d_\mathrm{tot}+r_\mathrm{T}-r_\mathrm{S})^2(d_\mathrm{tot}+r_\mathrm{T}+r_\mathrm{S})^2},
\end{eqnarray*}
where $A$ is the (distance-dependent) Hamaker coefficient of gold. As we could not find convincing values for this quantity, we take $A \approx 5 \cdot 10^{-19}$ as an upper bound, which is two times higher than the highest commonly found values for most materials \cite{Vergstrom1997Hamaker}. 

The surface separation in the proposed setup is much higher than typical interaction distances of forces emerging from dipole fluctuations (i.e. London-Van der Waals, Casimir-Polder and Casimir forces\footnote{Insights into how these forces are related are given in \cite{Rodriguez2011Casimir,Genet2004VacuumFluctuations}.}). In order to be able to exclude that any of the aforementioned effects contributes a noticeable signal, we also take into account an expression for the Casimir force, which we find to be \cite{Teo2012Casimir}
\begin{eqnarray*}
F_\mathrm{Cas} =  \frac{3 k_B T}{32}\frac{r_\mathrm{T} r_\mathrm{S}}{r_\mathrm{T} + r_\mathrm{S}}   (d_\mathrm{tot}-r_\mathrm{T}-r_\mathrm{S})^{-2} \zeta(3) \qquad \mbox{for} \qquad d_\mathrm{tot} \frac{k_B T}{\hbar c} \gg 1
\end{eqnarray*}
with the Riemanian zeta function $\zeta(z)$ and temperature $T$. 

For the effect of momentum transfer by residual gas molecules we estimate that an upper bound is given by
\begin{eqnarray*}
	| F_\mathrm{gas} | < \frac{r_\mathrm{T}^2 r_\mathrm{S}^2}{d_\mathrm{tot}^2} \pi P \sqrt{\frac{3 m_\mathrm{air}}{k_B T }}   | \dot d_\mathrm{tot} |
\end{eqnarray*}
where $P$ is the pressure, $m_\mathrm{air}$ is the molecular mass of air and $\dot x_\mathrm{S}$ is the source drive speed. An expression for the collision rate of air molecules onto the source mass is given in \cite{Chang2010LevitatedNanosphere}.

For the effect of patch potentials we estimate an upper bound as
\begin{eqnarray*}
	| F_\mathrm{patch} | < A_\mathrm{eff} \frac{\epsilon_0 V_\mathrm{rms}^2}{k_\mathrm{max}^2-k_\mathrm{min}^2} \frac{\partial}{\partial d_\mathrm{tot}}\int_{k_\mathrm{min}}^{k_\mathrm{max}}  \frac{k^2 \exp(-k (d_\mathrm{tot}-r_\mathrm{T}-r_\mathrm{S}))}{\sinh(k (d_\mathrm{tot}-r_\mathrm{T}-r_\mathrm{S}))} \rmd k
\end{eqnarray*}
which is an expression from \cite{Forces2003Speake,SurfaceContact2010Kim} for the assumption that the surface patch potential correlations are constant in a certain wave number range $k_\mathrm{min}<k<k_\mathrm{max}$. In \cite{Forces2003Speake} the central wave number is taken such that the integrand is maximal for the given distance, and a width of one decade around that band is chosen to set the integration boundaries. However, for our system that number would correspond to a wavelength larger than the source mass. Therefore we take the source mass size to be the maximum wavelength and the smallest wavelength to be one decade smaller. We apply the commonly used value $V_\mathrm{rms}\approx 90 \unit{mV}$ for gold and choose the cross section of the smaller of both spheres as the effective area. This is a very rough approximation of the situation as the above expression is only valid for the geometry of two planes, but as the interaction area could not be larger than the smaller of both cross-sections, this yields a reasonable upper bound to the strength of the effect.

\section{Test mass finite element simulations\label{app:fem}}
\begin{figure}[!htb]
	\centering
	\includegraphics{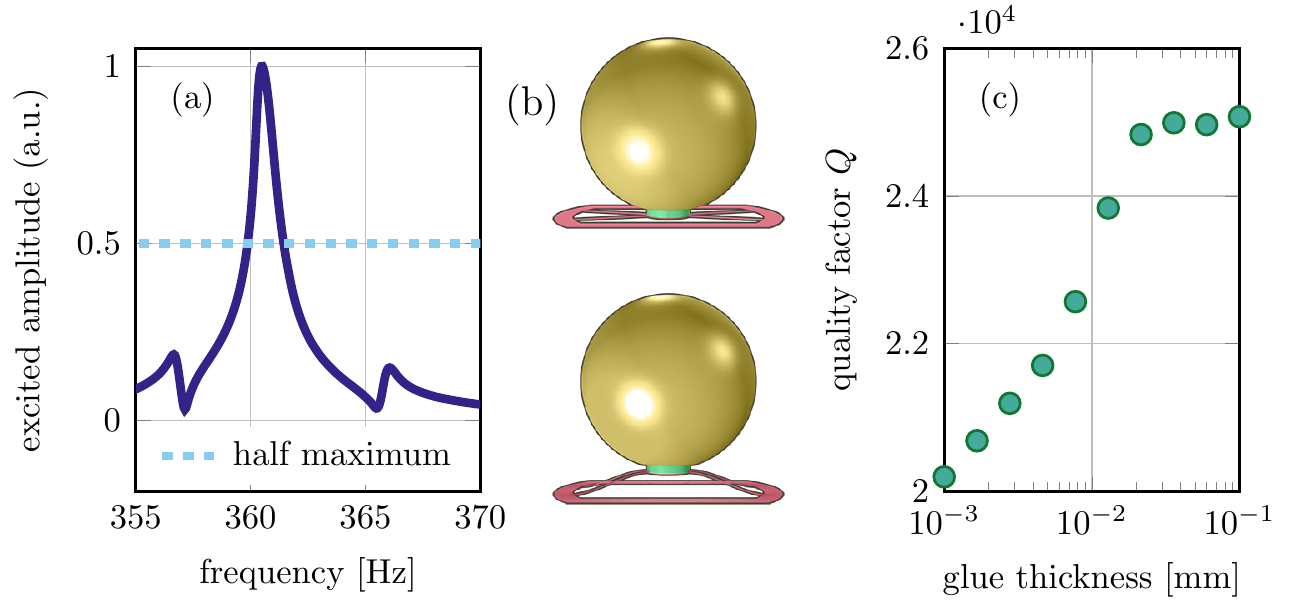}
	\caption{(a) Spectral forward transmission (S21 gain/loss) of a $2\unit{mm}$ gold sphere mounted on opposite poles. (b) Basic cantilever geometry with four arms and mirror pad, adhesive layer and $2\unit{mm}$ sphere (top), and the relevant mode with color-coded energy density (bottom). (c) Mechanical qualify factor as a function of the thickness of the adhesive layer as obtained by the FEM simulation.\label{fig:fem}}
\end{figure}

For devices that are not geometrically limited in mechanical quality but due to internal losses, the effective $Q$ value of a compound system can be computed as $Q^{-1} = U^{-1} \sum_i Q_i^{-1} U_i$ \cite{Harry2002ThermalNoise}. Here, $U$ is maximum elastic energy of the excited mode, $U_i$ is the part of the energy stored in the $i$th component and $Q_i$ is its quality factor. Specifically in our case,
\begin{eqnarray}
Q = U \left(  \frac{U_\mathrm{substrate}}{Q_\mathrm{substrate}} + \frac{U_\mathrm{adhesive}}{Q_\mathrm{adhesive}} + \frac{U_\mathrm{mass}}{Q_\mathrm{mass}}  \right)^{-1} , \label{eq:qcompound}
\end{eqnarray}
where \emph{substrate} denotes the cantilever material. With finite element method (FEM) eigenmode simulations we can tune a cantilever geometry to roughly oscillate at $50\unit{Hz}$ with the out-of-plane center of mass mode. We may then integrate the energy density in the deflected state for all individual components of the system for any given mode, allowing us to calculate $Q$ using the above expression. The individual $Q$-values were gathered as follows: For the adhesive, a $Q_\mathrm{adhesive}$ value of ``more than 300'' at room temperature has been reported in \cite{Schediwy2005HighQBonding}. As a substrate we assume AlGaAs, similar to the devices in \cite{Cole2012CrystallineMirrors}, with a room temperature $Q_\mathrm{substrate}\approx30{,}000$. As we could not find a convincing value for the mechanical quality of bulk gold, we performed a forward transmission measurement, also referred to as S21 gain/loss, of a $2\unit{mm}$ gold sphere, \fref{fig:fem}\textbf{(a)}, which was implemented by piezomechanical excitation and readout. With the half maximum width $\Delta f = \gamma / \pi \approx 1.6\unit{Hz}$ and central frequency $f_0 \approx 360.5\unit{Hz}$, we can estimate the mechanical quality as $Q=2 \pi f_0 / \gamma \approx 456$. However, as we do not expect more than order-of-magnitude precision out of this measurement,  we use a safer value of $Q_\mathrm{mass}=100$ for the computation. 
The geometry we chose for the FEM simulation is a 4-arm-cantilever with a central mirror pad, \fref{fig:fem}\textbf{(b)}. The AlGaAs substrate has a thickness of $7\unit{\mu m}$ and is rigidly clamped on the outer boundaries. The length of the arms is roughly $1\unit{mm}$. This yields a center of mass out-of-plane mode frequency of roughly $53\unit{Hz}$. Applying \eref{eq:qcompound} to the results obtained from the FEM simulations we estimate the effective mechanical quality of the compound system as a function of the adhesive thickness, which is shown in \fref{fig:fem}\textbf{(c)}. As the mechanical quality decreases when the adhesive layer thickness is smaller than $20\unit{\mu m}$, it will be necessary to apply a suitable minimal amount when assembling the actual structures.

\section{Brownian force noise from gas impacts\label{app:gasimp}}
To ensure that the mechanical quality will be limited by internal losses, we estimate the effect of additional Brownian force noise from residual air molecule impacts. As significant effects have been observed for larger test masses in constraint volumes \cite{Cavalleri2009Increased}, one may ask if partially constraining the flow of gas by a membrane between the two masses will increase residual gas damping to a relevant degree. As a worst-case scenario, we consider not a sphere, but a cylinder-shaped test mass with one flat surface parallel to a plane, as this allows the air molecules to bounce between both surfaces multiple times. Adjusting for conventions, an analytic expression for the damping rate $\gamma_\mathrm{air}$ through air collisions in the molecular regime (i.e. with the mean free path of molecules being much larger than all relevant spatial dimensions) can be found in \cite{Dolesi2011Brownian}:
\begin{eqnarray*}
\fl	\gamma_\mathrm{air}(P) = \left( \frac{\pi m_\mathrm{air}}{k_B T} \right)^{1/2} \frac{\pi r_\mathrm{T}^2}{m} \left[ \underbrace{\sqrt{8} \left(  1+\frac{h_\mathrm{T}}{2r_\mathrm{T}} + \frac{\pi}{4}  \right)}_{\mbox{free damping}}+ \underbrace{\frac{r_\mathrm{T}^2}{\sqrt{2} d_\mathrm{mem}^2 \ln (1+r_\mathrm{T}^2/d_\mathrm{mem}^2)}}_{\mbox{proximity damping}}  \right]  P
\end{eqnarray*}
with a test-mass radius~$r_\mathrm{T}$, a test-mass height~$h_\mathrm{T}$ and a separation~$d_\mathrm{mem}$ between the test mass surface and the membrane. Choosing $r_\mathrm{T}=1\unit{mm}$, $h_\mathrm{T}=2\unit{mm}$ and $d_\mathrm{mem}=1\unit{\mu m}$ and with a background pressure of $P=10^{-8}\unit{hPa}$, $\gamma_\mathrm{air}$ can only exceed the damping rate given by internal mechanical losses, $\gamma = \omega_0/Q$, if the mechanical quality~$Q$ is larger than $10^7$. In this case one would have to decrease the background pressure or increase the distance between test mass and membrane; both of which can easily be done as our assumed parameters are conservative. 

\newpage

{\footnotesize

\bibliographystyle{myunsrtnat}
\bibliography{bib,patent}
}

\end{document}